\documentstyle[amssymb,aps,epsfig,prl,twocolumn]{revtex}

\begin{document}
\title{Nuclear ferromagnetism induced Fulde-Ferell-Larkin-Ovchinnikov state}
\author{A. M. Dyugaev$^{1,2}$,I.D. Vagner$^{2,3}$ and P. Wyder$^{2}$}
\address{$^{1}${\it L.D. Landau Institute for Theoretical Physics, 7334}\\
Moscow, Russia.\\
{\it \ }$^{2}${\it Grenoble High Magnetic Field}\\
Laboratory,Max-Planck-Institut f\"{u}r Festk\"{o}rperforschung and \\
{\it CNRS, \ BP 166, 38042,Grenoble Cedex 09, France.}\\
$^{3}${\it \ Department of Communication Engineering, Holon Academic}\\
Institute of Technology, \\
{\it 52 Golomb Str., P.O.B. 305, Holon 58102, Israel.}}
\maketitle

\begin{abstract}
ABSTRACT

We present a theoretical study of the influence of the nuclear
ferromagnetism on superconductivity in the presence of the electron-nuclear
spin interaction. It is demonstrated that in some metals, e.g. Rh, W, the
BCS condensate imbedded in a matrix of ferromagneticaly ordered nuclear
spins should manifest the FFLO (Fulde-Ferel-Larkin-Ovchinniov) state. We
outline that the optimal experimental conditions for observation of FFLO
could be achieved by creation, via adiabatic nuclear demagnetization, of the
negative nuclear spin temperatures. In this case the nuclear polarization
points in the opposite to the external magnetic field direction and the
electromagnetic part of the nuclear spin magnetization compensates the
external magnetic field, while the exchange part creates the nonhomogeneous
superconducting order parameter.

PACS:
\end{abstract}

The problem of coexistence of the superconducting and magnetic ordering , in
spite of its long history \cite{Ginzb57}, is still among the enigmas of the
modern condensed matter physics. Most of the theoretical and experimental
efforts were devoted to studies of the coexistence of the electron
ferromagnetism and superconductivity.

Recently a growing interest to the physics of the superconducting state in
the presence of the nuclear magnetism has appeared both theoretically \cite
{DVW96} -\cite{Sonin98} and experimentally \cite{RHP97}-\cite{RhKTLJRN01}.
In \cite{RHP97} the magnetic critical field $H_{c}(T)$ of a metallic
compound $\ AuIn_{2}$ with $T_{ce}=0.207K$ was studied up to the
temperatures lower than the temperature of the nuclear spin ferromagnetic
ordering $T_{cn}$ . It was observed that the magnetic critical field $%
H_{c0}=14.5$ G has fallen by almost a factor of two at $T<T_{cn}=35\mu K$.
The possibility of such a reduction of $H_{c}(T)$ by the nuclear
ferromagnetism was outlined in \cite{DVW96}. Later on it was theoretically
considered in more details \cite{DVW97,KBB97,Sonin98}.

The critical magnetic field in the presence of magnetic moment of nuclear
spins is defined in the first approximation by the standard expression \cite
{Ginzb57}

\begin{equation}
H_{c}\left( T\right) =H_{co}\left( T\right) -4\pi \left( 1-n\right)
M_{n}(H_{c})  \label{Hc(T)a}
\end{equation}

Here $M_{n}$ is the nuclear magnetization and $n$ is the demagnetizing
factor, depending on the sample form. It follows from the Eq. $\left( \ref
{Hc(T)a}\right) $ that the difference between $H_{c}\left( T\right) $ and $%
H_{co}\left( T\right) $ is maximal in cylindric samples $\left( n=0\right) $
and vanishing in thin plate samples $\left( n=1\right) $.

In the very low temperature limit, $T<<$ $T_{ce}$ the temperature dependence
of the magnetic critical field is quite weak 
\begin{equation}
H_{co}\left( T\right) =H_{co}\left( 1-\frac{T^{2}}{T_{ce}^{2}}\right)
\label{Hco(T)a}
\end{equation}

The difference $H_{c}-H_{co}$ therefore depends mostly on the initial
conditions at the adiabatic nuclear demagnetizing procedure: the applied
magnetic field $H_{i}$ and magnetic nuclear spin moment $M_{n}\left(
H_{i}\right) $. In the limit when the initial magnetic field $H_{i}$ is
sufficiently larger than the local nuclear field $h$, \ $M_{n}\left(
H_{i}\right) $ is defined by the following expression

\begin{equation}
M_{n}(H_{i})=M_{no}B_{s}(x)  \label{Mn(Hi)a}
\end{equation}
where $x=$ $\frac{\mu _{n}H_{i}}{sT}$, $B_{s}(x)$ is the Brillouin function
and $s$ is the nuclear spin, $M_{no}$ is the saturation value of the nuclear
spin magnetisation $M_{no}=\mu _{n}n_{n\text{ }}$, $\mu _{n}$ and $n_{n\text{
}}$ are the nuclear magnetic moment and the nuclear spin density,
respectively.

The final nuclear magnetization $M_{n}\left( H_{f}\right) $ in the field $%
H_{f}$ could be found from \cite{Hebel63} 
\begin{equation}
M_{n}(H_{f})=M_{n}(H_{i})\frac{H_{f}}{\sqrt{h^{2}+H_{f}^{2}}}
\label{Mn(Hf)a}
\end{equation}
Eq-s $\left( \ref{Hc(T)a}\right) -\left( \ref{Mn(Hf)a}\right) $define the
influence of the ''electromagnetic'' part of the nuclear spin ordering on
the superconducting critical field $H_{c}$.

Recently the shift of $H_{c}\left( T\right) $ as a function of $M_{n}$ was
measured in the several metals: $Al$ \cite{SHP98}, $Sn$ \cite{Hermansd00}, $%
In$ \cite{InHT01}, and $Rh$ \cite{RhKTLJRN01}. It was found that the
experimental data for $Al$ and $In$ do not fit the Eq. $\left( \ref{Hc(T)a}%
\right) $, i.e. the shift in the critical field is not linear in $M_{n}$
when $M_{n}$ is approaching to its saturated value $M_{no}$ .

In \cite{DVW96,DVW97} we have suggested that apart of the influence of the
''electromagnetic'' part of the polarized nuclear spins on the
superconducting order, the hyperfine coupling between the nuclear spins and
conduction electron spins may play a crucial role on the coexistence between
superconducting state and nuclear ferromagnetism. It was shown also \cite
{DVW01} that creation of the negative nuclear spin temperatures, $NNST$, may
result in enhancement of the superconducting ordering. In this Letter we
demonstrate that the hyperfine part of the nuclear-spin-electron interaction
may result, in some metals, in appearance of the nonuniform superconducting
order parameter, the so called Fulde-Ferell-Larkin-Ovchinnikov state (FFLO) 
\cite{FF64,LO65}. It follows, that the range of parameters where FFLO can
exist is much wider under the conditions of $NNST$.

The FFLO state was thought originally to take place in superconductors with
magnetically ordered magnetic impurities \cite{FF64,LO65}. The main
difficulty, hovewer, in the observation of the FFLO in this case is in
simultaneous action of the ''electromagnetic'' and ''exchange'' parts of the
magnetic impurities on the superconducting order. In most of the known
superconductors the ''electromagnetic'' part is destroying the
superconducting order before the ''exchange'' part modify the BCS condensate
to a nonuniform FFLO state.

The situation may change drastically in the case of the nuclear spin
ferromagnetic ordering. Indeed, the nuclear magnetic moment $\mu _{n}=\frac{%
\hbar e}{M_{i}c},$ is at least three orders of magnitude smaller than the
electron Bohr magneton $\mu _{e}=\frac{\hbar e}{m_{o}c},$so that the
''electromagnetic'' part of the nuclear spin fields is quite low, compared
to that of the magnetic impurities. By the other hand the ''exchange'' part
is strongly dependent on the nuclear charge $Z$. In what follows, we define
the conditions and materials where the interplay between these two
contributions can be in favor for the ''exchange'' part, thus providing the
necessary conditions for appearance of the FFLO.

As in the case of the magnetic impurities \cite{FF64,LO65}, the polarized
nuclear spins also remove the spin degeneracy of conduction electrons

\begin{equation}
E_{\pm }=\sqrt{\Delta ^{2}+\left( p-p_{F}\right) ^{2}v_{F}^{2}}\pm J
\label{E+-a}
\end{equation}
where $\Delta $ is the gap in the electron spectrum, $p_{F}$ and $v_{F}$ the
electron Fermi momentum velocity. The parameter $J\equiv \mu _{e}H_{hyp}$
defines the Zeeman splitting of the electron spins, due to the hyperfine
interaction between the polarised nuclear spins and the conduction electron
spins. We note that $H_{hyp}$ is proportional to the magnetization $M_{n}$
produced by the nuclear spins \cite{DVW97} and $H_{hyp}^{\max }\equiv H_{no}$%
, the maximal value of $H_{hyp}$, is achieved when $M_{n}=M_{n}^{\max }$ $%
\equiv M_{no}$.

By introducing the reduced nuclear magnetization $M_{n}^{\ast }\equiv \frac{%
M_{n}}{M_{no}}$, we can write $H_{n}=H_{no}M_{n}^{\ast };$ \ $%
J=J_{o}M_{n}^{\ast }$where $J_{o}\equiv $ $\mu _{e}H_{no}$. $H_{no}$ and $%
J_{o}$could be found from the expression \cite{Abragam61} 
\begin{equation}
H_{no}=\frac{\kappa }{\chi }M_{no}  \label{Hnob}
\end{equation}
where $\kappa $ is the Knight shift constant and $\chi $ is the conduction
electron paramagnetic susceptibility. $\kappa $ was measured for most of
metals and can be as large as $10^{-2}$ for nuclei with large $Z$ since $%
\kappa \sim Z$ \cite{Abragam61}. At present $\chi $ is experimentally
defined only for $Li$ and $Na$ to be of order of $10^{-6}$.

Using the Fermi liquid theory, the electronic susceptiility is defined by
the relation $\chi =\chi _{o}\frac{1}{1+Z_{o}}$where $Z_{o}$ is the Fermi
liquid constant, $\chi _{o}=\mu _{e}^{2}\nu $ and $\nu $ is the density of
electronic states which could be defined from the low-temperature value of
the specific heat in a normal metal \cite{AGD}, $C\left( T\right) =\frac{\pi
^{2}}{3}\nu T$. In superconductors, hovewer, there exist an alternative
definition of $\nu $ via the gap $\Delta $ and the critical magnetic field $%
H_{co}$

\begin{equation}
\nu \Delta ^{2}=\frac{H_{co}^{2}\left( 0\right) }{2\pi };\text{ }\Delta
=1.76T_{ce}  \label{nuDel2}
\end{equation}
Unfortunately, the Fermi liquid constant $Z_{o}$ in a superconducting state
is not known.

We will, therefore, estimate $\chi _{o}$ using the Eq-s $\left( \ref{nuDel2}%
\right) $ and known experimental values for $H_{co}\left( 0\right) $ and $%
T_{ce}$. \ In Table 1 we present the values of $\chi _{o}$ and nuclear spin
field $H_{no}$, Eq. $\left( \ref{Hnob}\right) $. For most of superconductors
the hyperfine nuclear spin field H$_{no}$ is up to four orders of magnitude
larger than the magnetic moment of the polarized nuclei $M_{no}$: $\frac{%
\kappa }{\chi }\sim 10^{4}$ . As it was discussed previously \cite{DVW01},
the nuclear field $H_{no}$ and the parameter $Z_{o}$ in superconductors
could be defined from experimental data on $H_{c}\left( T\right) $ and $%
H_{co}\left( T\right) $ using the expression

\begin{equation}
\left[ H_{c}\left( T\right) +4\pi \left( 1-n\right) M_{n}\right]
^{2}=H_{co}^{2}\left( T\right) -2\frac{J^{2}}{\Delta ^{2}}H_{co}^{2}\left(
0\right)  \label{12a}
\end{equation}

At low enough temperatures $H_{c}\left( T\right) \simeq H_{c}\left( T\right) 
$ and Eq. $\left( \ref{12a}\right) $ can be written in reduced variables in
the form

\begin{equation}
\left[ H_{c}^{\ast }+\left( 1-n\right) \varepsilon M_{n}^{\ast }\right]
^{2}=1-2\lambda ^{2}M_{n}^{\ast ^{2}}  \label{12b}
\end{equation}
where the reduced variables are defined as follows: $H_{c}^{\ast }=\frac{%
H_{c}}{H_{co}}$, $M_{n}^{\ast }\equiv \frac{M_{n}}{M_{no}}$, $\varepsilon =%
\frac{4\pi M_{no}}{H_{co}}$, $\lambda =\left( 1+Z_{o}\right) \lambda _{o}$
and $\lambda _{o}=$ $\varepsilon \frac{\Delta \kappa }{2\mu _{e}H_{co}}$.
The values of $\varepsilon $ and $\lambda _{o}$ for several superconductors
are given in the Table 1. The values of $M_{no}$ for all the known
superconductors are given in \cite{HRSP98}.

While analyzing the Eq. $\left( \ref{12b}\right) $ one should bear in mind
the possibility of nuclear spin system having an either positive or negative
temperature, as it was outlined by us in \cite{DVW01}. \ 

Consider first the {\bf positive nuclear spin temperature} $T_{n}>0$. In
this case, the nuclear magnetization is directed along the applied field and
Eq. $\left( \ref{12b}\right) $ has a single valued solution

\begin{equation}
H_{c}^{\ast }=\sqrt{1-2\lambda ^{2}M_{n}^{\ast ^{2}}}-\left( 1-n\right)
\varepsilon M_{n}^{\ast }  \label{Hc*b}
\end{equation}
Two experimental metodiques of nuclear spin cooling should be considered:
the adiabatic demagnetization and the dynamic polarisation \cite
{Abragam61,SlichterBk}.

a) Nuclear polarization is created by the adiabatic demagnetization. In this
case

\begin{equation}
M_{n}^{\ast }=\frac{H_{c}^{\ast }}{\sqrt{h^{\ast 2}+H_{c}^{\ast 2}}}%
B_{S}\left( \frac{\mu _{n}H_{i}}{sT}\right) ;h^{\ast }=\frac{h}{H_{co}}
\label{Mn*b}
\end{equation}

It is easy to see that the solution $H_{c}^{\ast }=0$ do not satisfy the
Eq-s. $\left( \ref{Hc*b}\right) $ and $\left( \ref{Mn*b}\right) $
simulataneosly. This means that the nuclear spins polarized by adiabatic
demagnetization, while reducing $H_{c}$, do not completly destroy the
superconductivity.

b) Nuclear polarization is created by the methods of dynamic polarisation.
In this case the Eq. $\left( \ref{Mn*b}\right) $ is not valid and the $%
H_{c}^{\ast }=0$ solution can be obtained at sutable choise of parameters.
Indeed, even in the case of a thin plate, $n=1$, the BCS superconducting
state vanishes at $\lambda M_{n}^{\ast }>\frac{1}{\sqrt{2}}=0.707$. \ Since $%
M_{n}^{\ast }<1$, a first order phase transition from the superconducting
phase to a normal one will occure in supperconductors with $\lambda >\frac{1%
}{\sqrt{2}}$.

In \cite{FF64,LO65} it was conjectured, hovewer, that even at $\lambda
M_{n}^{\ast }>\frac{1}{\sqrt{2}}$ a superconducting state with a
nonhomogeneous order parameter may appear, the long searched Fulde - Ferell
- Larkin - Ovchinnikov state (FFLO). In \cite{FF64,LO65} it was found that
the difference in the free energies of a normal and nonhomogeneous
superconducting states, at $J\lesssim J_{c}$, is: $f_{s}-f_{n}=-C_{o}^{2}\nu
\left( J_{c}-J\right) ^{2}$; $J_{c}\simeq 0.755$. Using the general
thermodynamic arguments \cite{AGD} the critical field $\overline{H}%
_{c}^{\ast }$ can be defined from an equation

\begin{equation}
4C_{o}^{2}\lambda ^{2}\left( M_{n}^{\ast }-M_{nc}^{\ast }\right) ^{2}=\left[ 
\overline{H}_{c}^{\ast }+\left( 1-n\right) \varepsilon M_{n}^{\ast }\right]
^{2}  \label{20x}
\end{equation}
Here $\overline{H}_{c}^{\ast }\equiv \frac{\overline{H}_{c}}{H_{co}}$, where 
$\overline{H}_{c}$ is the critcal field of the FFLO phase and $M_{nc}^{\ast }
$ is the critical value of the nuclear magnetization:$\lambda M_{n}^{\ast
}=0.755.$

In \cite{LO65} different forms of the nonhomogeneous order parameter were
considered. It was found there that the FFLO state with an order parameter $%
\Delta =\Delta _{o}\exp \left\{ i\overrightarrow{q}\cdot \overrightarrow{r}%
\right\} $ can exist in a narrow interval $0.707>\lambda M_{n}^{\ast }>0.755$
with a constant $C_{o}^{2}=0.44$ , see Fig.1, the solid line. It is
interesting to note that for an order parameter $\Delta =\Delta _{o}\cos
\left\{ \overrightarrow{q}\cdot \overrightarrow{r}\right\} $ the constant $%
C_{o}^{2}=7.35$ and the FFLO state may exist in a much wider interval of
values of $\lambda M_{n}^{\ast }$ , see Fig.1, the dashed line. One should
bear in mind, hovewer, that this wide interval is limited by the made above
assumption: $J\lesssim J_{c}$.

At positive nuclear temperatures, $T_{n}>0$ , \ the Eq. $\left( \ref{20x}%
\right) $ has one solution

\begin{equation}
\overline{H}_{c}^{\ast }=2C_{o}\lambda \left( M_{n}^{\ast }-M_{nc}^{\ast
}\right) -\left( 1-n\right) \varepsilon M_{n}^{\ast }  \label{21x}
\end{equation}

Eq. $\left( \ref{Hc*b}\right) $ and Eq. $\left( \ref{21x}\right) $ define
the $H_{c}^{\ast }$,$\overline{H}_{c}^{\ast }$ vs $M_{n}^{\ast }$ phase
diagram for two superconducting and one normal metal phases, see Fig. 1.

Among the possible candidates for the observation of nuclear spin
polarisation induced FFLO state are $Rh$ and $W$. $Rh$ was studied in
details in \cite{RhKTLJRN01}. It has rather high $\varepsilon =8.2$ and $%
\lambda _{o}=2.8$. The transition to a superconducting phase with a
nonuniform order parameter is possible when the reduced magnetization $%
M_{n}^{\ast }>0.25$. The local field $h$ is hovewer also rather high: $%
h^{\ast }=\frac{h}{H_{co}}=6.9$ ,\cite{RhKTLJRN01}. This smakes it difficult
to get sufficiently high values of $M_{n}^{\ast }$ using the adiabatic
demagnetization method.

\bigskip

\begin{tabular}{cccccc}
Elements & H$_{co}(G)$ & H$_{no}(G)$ & $\chi _{o}10^{6}$ & $\varepsilon $ & $%
\lambda _{o}$ \\ 
AuIn$_{2}$ & 14.5 & 5720 & 1.14 & 0.7 & 1.06 \\ 
Al & 105 & 957 & 1.85 & 0.13 & 0.03 \\ 
Mo & 95 & 703 & 2.47 & 0.04 & 0.03 \\ 
Rh & 0.049 & 22 & 6.2 & 8.2 & 2.8 \\ 
Cd & 30 & 615 & 0.8 & 0.06 & 0.05 \\ 
Ta & 830 & 902 & 8.0 & 0.01 & 0.008 \\ 
W & 1.2 & 347 & 1.48 & 0.52 & 0.88 \\ 
Ir & 19 & 185 & 4.27 & 0.04 & 0.05 \\ 
Tl & 171 & 3970 & 1.19 & 0.02 & 0.06
\end{tabular}

\bigskip

It follows from the Eq. $\left( \ref{Hc*b}\right) $ and Eq. $\left( \ref
{Mn*b}\right) $ that in superconductors with $h^{\ast }>>1$

\begin{equation}
H_{c}^{\ast }\simeq \left[ \left( 1+\frac{\varepsilon }{h^{\ast }}\left(
1-n\right) B_{S}\right) ^{2}+2\lambda ^{2}\frac{B_{S}}{h^{\ast 2}}\right] ^{-%
\frac{1}{2}}  \label{Hc*d}
\end{equation}

For $W$, with $\varepsilon =0.52$ and $\lambda _{o}=0.88$, the nuclear spin
relaxation time is very long (the Korringa's const. $\kappa =39Ks$). This
makes it feasible, in $W$, to get nuclear temperatures $T>>T_{n}$ by
adiabatic demagnetisation.

At the {\bf negative nuclear spin temperatures}, ($NNST$), $T_{n}<0$, the
nuclear magnetisation $M_{n}$ points in the direction opposite to the
applied field $H$ \cite{DVW01}, and the direction of $M_{n}$ can be reversed
by a well known method of fast reversal of the external field \cite{OL97}.
In \cite{DVW01} we have shown that the $NNST$ may strongly enhance the
superconducting ordering. For the metals like $Be$, TiH$_{2.07}$ , for
example, the critical field $H_{c}$ is order of magnitude higher than $%
H_{co} $ at saturation of $M_{n}$.

Let us show now that the $NNST$ stimulate the appearrance of the FFLO phase.
\ As it was shown earlier in this paper, at $T_{n}>0$, \ the Eq. $\left( \ref
{12b}\right) $ and Eq. $\left( \ref{20x}\right) $ do not posses a solution
with low $H_{c}^{\ast }$, where the FFLO state should appear, see Fig. 1. At 
$T_{n}<0$, hovewer, the Eq. $\left( \ref{12b}\right) $ and Eq. $\left( \ref
{20x}\right) $ have two solutions for $H_{c}^{\ast }$ and $\overline{H}%
_{c}^{\ast }$

\begin{equation}
H_{c}^{\ast }=\left( 1-n\right) \varepsilon M_{n}^{\ast }\pm \sqrt{%
1-2\lambda ^{2}M_{n}^{\ast ^{2}}}  \label{Hc*a}
\end{equation}

\begin{equation}
\overline{H}_{c}^{\ast }=\left( 1-n\right) \varepsilon M_{n}^{\ast }\pm
2C_{o}\lambda \left| M_{n}^{\ast }-M_{nc}^{\ast }\right|  \label{Hc*c}
\end{equation}

The phase diagram, in this case, is presented in Fig. 2. \ In this limit the
FFLO\ phase can be realized in the high field region which makes it
observation much easier than in the case of the positive nuclear
temperatures, Fig. 1. We outline here that the application of a rather
strong magnetic field at $T_{n}<0$ will not destroy, as is the case at $%
T_{n}>0$, but rather stabilize the FFLO state.

Let us discuss now the experimental feasibility of our model. The most
suitable experimental conditions for observation of the FFLO state would be
the case when the nuclear spin ordering appears first and than by lowering
the external magnetic field the superconducting state is established \cite
{DVW97}. In this limit the nuclear spin ordering is not modified
substantially by superconductivity, since the nuclear spin relaxation times,
at micro-kelvin temperatures, are extremely long. This can be achieved by
adiabatic nuclear demagnetization at nuclear temperatures $T_{n}$ much
higher than $T_{cn}$, the temperature of spontaneous nuclear ordering. In
this case the sample is in a monodomain state. The superconducting ordering
starts at the electron temperature $T$ in the interval $T_{n}<<$ $T<<$ $%
T_{ce}$ \ and critical magnetic field $H_{c}(T),$which is different from $%
H_{co}(T)$, the critical magnetic field in the absence of the nuclear spin
ordering.

It was shown above the FFLO state could be easier obtained under the
conditions of negative nuclear spin temperature, since the electromagnetic
part of the nuclear spin field reduces the influence of the external field
and the exchange part will act to create the nonhomogeneous superconducting
ordering. Experimentaly the $NNST$ was achieved in single crystal $Rh$ \cite
{RhKTLJRN01} employing the nuclear demagnetizing techniques. Hovewer, no
study of possible superconducting order was performed. It follows, from the
considerations presented above, that the FFLO state may appear under similar
to \cite{RhKTLJRN01} conditions at critical fields $H>H_{co}$

We acknowledge stimulating discussions with A.G.M.\ Jansen, W. Joss, V.
Mineev and Yu. Ovchinnikov. This work was supported by the Russian
Foundation for Basic Research, project no. 00-02-17729, the Israeli Academy
of Sciences and by the European Commission, IST-2000-29686. A.D. is grateful
to G. Martinez for the hospitality at GHMFL during the work on this paper.

{\bf Figure Captions.}

Fig. 1. The phase diagram for the Fulde-Ferell-Larkin-Ovchinnikov state for
two types of nonehomogeneous order prameter. a) an order parameter $\Delta
=\Delta _{o}\exp \left\{ i\overrightarrow{q}\cdot \overrightarrow{r}\right\} 
$ and b) $\Delta =\Delta _{o}\cos \left\{ \overrightarrow{q}\cdot 
\overrightarrow{r}\right\} $. Positive nuclear temperatures, $T_{n}>0$, and
the demagnetising factor $n=1$.

Fig. 2. The phase diagram for the Fulde-Ferell-Larkin-Ovchinnikov state in $%
Rh$ with $\varepsilon =8.2$, $\lambda =2.8$ and the demagnetising factor $%
n=0 $, at negative nuclear temperatures, $T_{n}<0$, for two types of
nonehomogeneous order prameter: a) $\Delta =\Delta _{o}\exp \left\{ i%
\overrightarrow{q}\cdot \overrightarrow{r}\right\} $ and b) $\Delta =\Delta
_{o}\cos \left\{ \overrightarrow{q}\cdot \overrightarrow{r}\right\} $.


\begin{references}
\bibitem{Ginzb57}  V.L. Ginzburg, Sov. Phys. JETP, {\bf 4}, 153 (1957).

\bibitem{DVW96}  A.M. Dyugaev, I.D. Vagner and P. Wyder, JETP-Lett., {\bf 64,%
} 207 (1996).

\bibitem{DVW97}  A.M. Dyugaev, I.D. Vagner and P. Wyder, JETP-Lett., {\bf 65}%
,{\bf \ }810 (1997).

\bibitem{DVW01}  A.M. Dyugaev, I.D. Vagner and P. Wyder, JETP-Lett., {\bf 73}%
,{\bf \ }411 (2001).

\bibitem{KBB97}  M.L. Kulic, A.I. Buzdin and L.N. Bulaevskii, Phys. Rev. B, 
{\bf 56}, R11415 (1997).

\bibitem{Sonin98}  E.B. Sonin, J. Low Temp. Phys.,{\bf 110}, 411 (1998).

\bibitem{RHP97}  S. Rehman, T. Herrmannsd\"{o}rfer, and F.Pobel, Phys. Rev.
Lett., {\bf 78}, 1122 (1997).

\bibitem{SHP98}  M. Seibold, T. Herrmannsd\"{o}rfer, and F.Pobel, J. Low
Temp. Phys.,{\bf 110}, 363 (1998).

\bibitem{Hermansd00}  T. Herrmannsd\"{o}rfer, Physica B{\bf 280}, 368 (2000).

\bibitem{HRSP98}  T. Herrmannsd\"{o}rfer, S. Rehman, M. Seibold, and
F.Pobel, J. Low Temp. Phys.,{\bf 110}, 405 (1998).

\bibitem{InHT01}  T. Herrmannsd\"{o}rfer and D. Tayurskii, J. Low Temp. Phys.%
{\bf 124},, Nos. 1/2, 257 (2001).

\bibitem{RhKTLJRN01}  T.A. Knuuttila, J.T. Tuoriniemi, K. Lefmann, K.I.
Juntunen, F.B. Rasmussen, K.K. Nummila, J. Low Temp. Physics, {\bf 123},
Nos. 1/2, p.65 (2001).

\bibitem{Hebel63}  L.C. Hebel, Solid State Phys., {\bf 15}, 409 (1963).

\bibitem{FF64}  P. Fulde, R.A. Ferrel, Phys. Rev. {\bf 135}, 1550 (1964).

\bibitem{LO65}  A.I. Larkin, Yu.N. Ovchinnikov, J. Exp. Theor. Phys., {\bf 47%
}, 1136 (1964) (Sov. Phys.,- JETP, {\bf 20}, 762 (1965)).

\bibitem{Abragam61}  A.A.Abragam, ''{\it The principles of Nuclear Magnetism}%
'' (Oxford Univ. Press, London and New York 1961);J. Winter, {\it Magnetic
Resonance in Metals}, (Oxford, 1971).

\bibitem{AGD}  A.A. Abrikosov, L.P. Gor'kov and I.E. Dzaloshinski, {\it %
Methods of Quantum Field Theory in Statistical Physics ,}(Pergamon,
Elmsford, N.Y. 1965).

\bibitem{SlichterBk}  C.P. Slichter, {\it Principles of Magnetic Resonance},
Third Edition (Springer-Verlag, Berlin, 1990).

\bibitem{OL97}  A.S. Oja and O.V. Lounasmaa, Rev. Mod Phys., {\bf 69}, 1
(1997).
\end{references}
\end{document}